\documentclass[fleqn,10pt]{wlscirep}
\usepackage[utf8]{inputenc}
\usepackage[T1]{fontenc}
\usepackage{bm}

\title{Controlling superconductivity of CeIrIn$_5$ microstructures by substrate selection}

\author[1,*,\dag]{Maarten R. van Delft}
\author[2,3]{Maja D. Bachmann}
\author[1]{Carsten Putzke}
\author[1]{Chunyu Guo}
\author[1]{Joshua A. W. Straquadine}
\author[4]{Eric D. Bauer}
\author[4]{Filip Ronning}
\author[1,*]{Philip J. W. Moll}
\affil[1]{Laboratory of Quantum Materials (QMAT), Institute of Materials (IMX), École Polytechnique Fédérale de Lausanne (EPFL), 1015 Lausanne, Switzerland}
\affil[2]{Max Planck Institute for Chemical Physics of Solids, 01187 Dresden, Germany}
\affil[3]{Scottish Universities Physics Alliance, School of Physics and Astronomy, University of St. Andrews, St. Andrews KY16 9SS, U.K.}
\affil[4]{Los Alamos National Laboratory, Los Alamos, NM 87545, USA.}

\affil[*]{e-mail: maarten.vandelft@ru.nl; philip.moll@epfl.ch}

\affil[$\dag$]{Present address: High Field Magnet Laboratory (HFML - EMFL), Radboud University, Toernooiveld 7, 6525 ED Nijmegen, The Netherlands}

\begin{abstract}
Superconductor/metal interfaces are usually fabricated in heterostructures that join these dissimilar materials. A conceptually different approach has recently exploited the strain sensitivity of heavy-fermion superconductors, selectively transforming regions of the crystal into the metallic state by strain gradients. The strain is generated by differential thermal contraction between the sample and the substrate. Here, we present an improved finite-element model that reliably predicts the superconducting transition temperature in CeIrIn$_5$ even in complex structures. Different substrates are employed to tailor the strain field into the desired shapes. Using this approach, both highly complex and strained as well as strain-free microstructures are fabricated to validate the model. This enables full control over the microscopic strain fields, and forms the basis for more advanced structuring of superconductors as in Josephson junctions.

\end{abstract}
\begin{document}

\flushbottom
\maketitle
\thispagestyle{empty}

\section*{Main text}
The local and selective transformation of materials properties forms the basis of electronics. In the transistor, for example, electric fields drive the transition between an insulator and a metallic conductor.  In correlated electron systems, phase transformations can be driven via relevant tuning parameters, such as strain. In thin film materials, strains caused by mismatch between substrate and film have been investigated for a long time \cite{Stoney1909}. Such mismatch strains can either be a nuisance, as a source of dislocations and materials incompatibility, or a blessing, leading to desirable band structure modifications. Each film thus presents a new challenge in finding the right substrate to achieve the desired level of strain \cite{Abadias2018,Engwall2016,Motazedian2021}.  In bulk crystals, however, strain and strain gradients are usually weak. When it is desirable to purposely induce strain, often an elaborate straining apparatus is required \cite{Kostylev2019,Hicks2014}.

Regardless of whether one is dealing with strain in a thin film or in a bulk crystal, the strain is usually applied globally to the entire film or crystal. Strategies do exist to induce controlled local strains in certain two-dimensional materials \cite{Foerster2020,Bollani2015}, but these do not translate well to three-dimensional solids, while a method to control strain locally could enable both novel types of basic science experiments as well as applications. Recently, controlled $T_c$ landscapes imprinted by static strain gradient fields of CeIrIn$_5$ have been demonstrated\cite{Bachmann2019}. The approach exploits the high sensitivity of the superconducting critical temperature, $T_c$, on directional strain, which translates strain gradients into $T_c$ gradients. Technically, this was achieved by shaping crystal microstructures via focused ion beam (FIB) and by generating complex strain patterns due to differential thermal contraction between the structure and the substrate.

Here we build on this work and demonstrate strongly improved control over the $T_c$ landscape. The main advances are a refinement of the finite element modeling of strain fields that is now able to predict detailed $T_c$ modifications, as well as an exploration of various substrates with differences in the thermal contraction. Consequently, desired $T_c$ landscapes can be reliably predicted and validated by experiment. We continue to focus on the heavy fermion superconductor CeIrIn$_5$ \cite{Petrovic2001}, but the principle is materials agnostic and can be transferred to any material. CeIrIn$_5$ is a convenient material, as its superconductivity is highly sensitive to uniaxial strain\cite{Dix2009,Oeschler2003}. In particular, the tetragonal compound reacts to strain in a directional way: while compression along the $a$-direction increases $T_c$, it is reduced by compression along the $c$-direction. 

\begin{figure}[h!]
\centering
\includegraphics[width=1\linewidth]{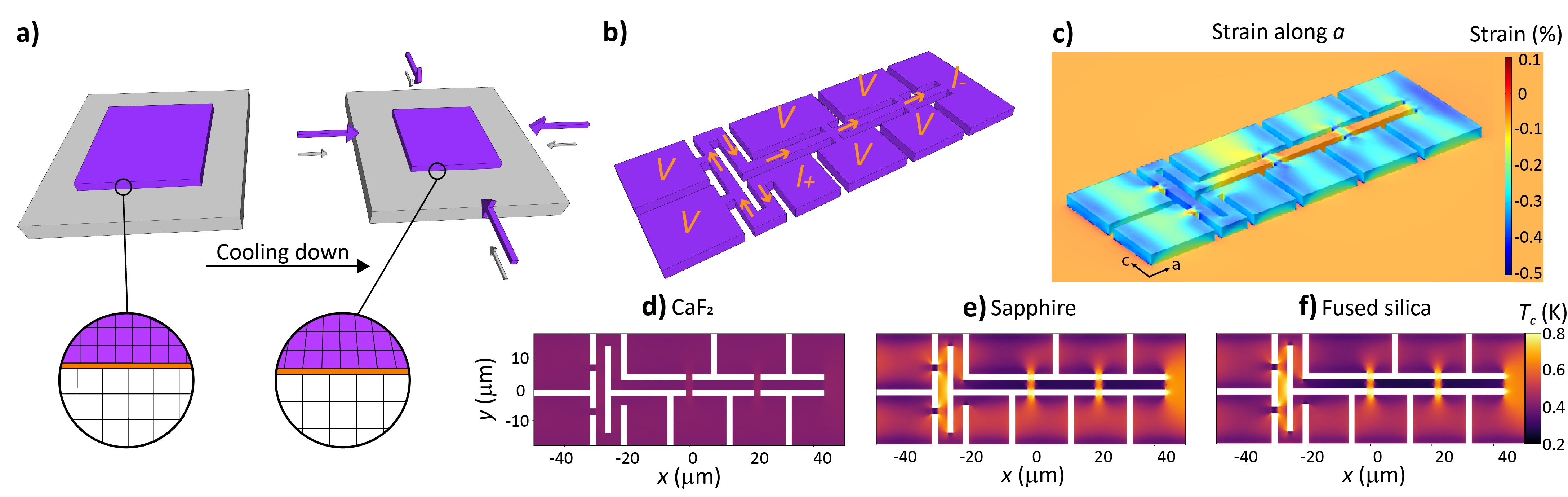}
\caption{(a) Illustration of the differential thermal contraction taking place in the CeIrIn$_5$/substrate devices. The crystal (purple) contracts more than the substrate (grey), leading to strain at the interface. The glue layer (brown) buffers any lattice mismatch. (b) Sketch of a T-shaped device geometry. The arrows indicate the current flow between the $I_+$ and $I_-$ pads. (c) Components of the strain tensor along the $a$-axis, shown on a T-shape device on a sapphire substrate. (d)-(f) Simulated critical temperatures, $T_c$, of T-shaped devices on different substrate materials: (d) CaF$_2$, (e) sapphire and (f) fused silica. $T_c$ is calculated in the horizontal center plane of the devices and the scale bar shown in (f) applies to (d), (e) and (f).}
\label{fig1}
\end{figure}

The origin of the strain in our system is illustrated in Fig~\ref{fig1}a. We fix a crystal slab, which we call a lamella, onto a substrate at room temperature using a thin layer of epoxy. As the sample is cooled to near-zero kelvin, the crystal thermally contracts more strongly than the substrate, causing strain. This strain is not uniform, but is distributed depending on the shape of the sample that sets the elastic boundary conditions. Using a FIB, we can modify this shape in order to prepare a device suitable for electrical measurements, with a desired strain pattern. Unlike in thin film interfaces, the glue layer buffers any lattice mismatch and hence we are not concerned with matching lattice parameters of the material and substrate. The relaxed structure is glued to the substrate at room temperature and cured at 140~$^{\circ}$C. Strain then arises from differences in thermal contraction when cooling down to low temperatures (see Fig~\ref{fig1}a). The fabrication method of these crystalline structures has been reported elsewhere\cite{Moll2018,Moll2015c}.

The magnitude and spatial distribution of the strain is modeled using finite element simulations implemented in the structural mechanics module of COMSOL multiphysics\cite{Bachmann2019} (see supplementary information for details).  Fig.~\ref{fig1}b-c displays a single-crystal T-shaped sample geometry. As indicated in Fig.~\ref{fig1}b, the central beam guides the current, while the lateral crystal pads serve as voltage contacts. Unavoidably, these contacts also transduce stress into the beam which then shapes the strain profile (shown in Fig.~\ref{fig1}c for strain along the a-axis).  This geometry is useful to determine the transport anisotropy in a crystal, as it allows resistance measurements along two different crystallographic axes, in this case the $a$- and $c$-axes of CeIrIn$_5$. The high level of complexity of such a structure is an ideal testbed for simulations, where the mechanical model is used to predict the components of the strain tensor along the $a$- and $c$-axes. We also calculate the other components, but these are small enough to be safely neglected (see supplementary Fig.~S1 for an overview of all components). 

The overall magnitude of the strain strongly depends on the choice of the substrate material and its thermal contraction. Previous work on CeIrIn$_5$\cite{Bachmann2019} and other materials\cite{Bachmann2019d} focused exclusively on sapphire as a substrate, while here we employ various materials to engineer the strain. Sapphire has a small coefficient of thermal expansion ($\alpha=\frac{1}{L}\frac{dL}{dT}$) and hence contracts little when cooled down. Combined with the strong contraction of CeIrIn$_5$, this leads to a sizeable strain. For higher strain applications, this can be further enhanced by using fused silica as substrate, a material with even lower $\alpha$ than sapphire ($\alpha_{\textrm{sapphire}}$=5.4$\times10^{-6}$~K$^{-1}$ versus $\alpha_{\textrm{fused silica}}$=0.5$\times10^{-6}$~K$^{-1}$  at room temperature\cite{Ekin2006,Hahn1972}). For low strain applications, on the other hand, it is desirable to use a substrate with $\alpha$ similar to that of the material under study; a convenient option for CeIrIn$_5$ is CaF$_2$ ($\alpha_{\textrm{CaF}_2}$=18.85$\times10^{-6}$~K$^{-1}$ at room temperature\cite{Batchelder1964}).

As expected, the choice of substrate has a significant effect on the strain and $T_c$ is strongly modulated within the structure (Fig.~\ref{fig1}d-f). The local $T_c$ is computed from the local ratio of the $c$ and $a$ lattice parameters, whose dependence has been previously obtained in uniaxial pressure experiments on macroscopic crystals\cite{Dix2009}. It is clear that significant variations in $T_c$ exist across any device on sapphire or fused silica, which will certainly affect measurements of these devices, for instance leading to multiple superconducting transitions seen in electrical measurements. If instead of strain engineering a measurement corresponding to the intrinsic properties of bulk CeIrIn$_5$ is desired, the use of a CaF$_2$ substrate is expected to reduce this variation to a minimum. As seen in Fig.~\ref{fig1}d, $T_c$ variations in this device on CaF$_2$ are considerably smaller, of the order of 20~mK.

For an optimal measurement of intrinsic CeIrIn$_5$ along well-defined crystallographic directions, the $T_c$ variations can be further reduced by adapting the geometry. Here it is important to note that larger $T_c$ variations arise near edges and corners. It is therefore important to avoid having those near the sections of the device that are being measured, i.e. between the contacts. Naturally, the contact connections themselves are points that transduce strain. We therefore further optimize the design to reduce $T_c$ variation near the contacts: the corners are rounded and the contact connections are softened by elongating and thinning them. 

\begin{figure}[ht]
\centering
\includegraphics[width=0.8\linewidth]{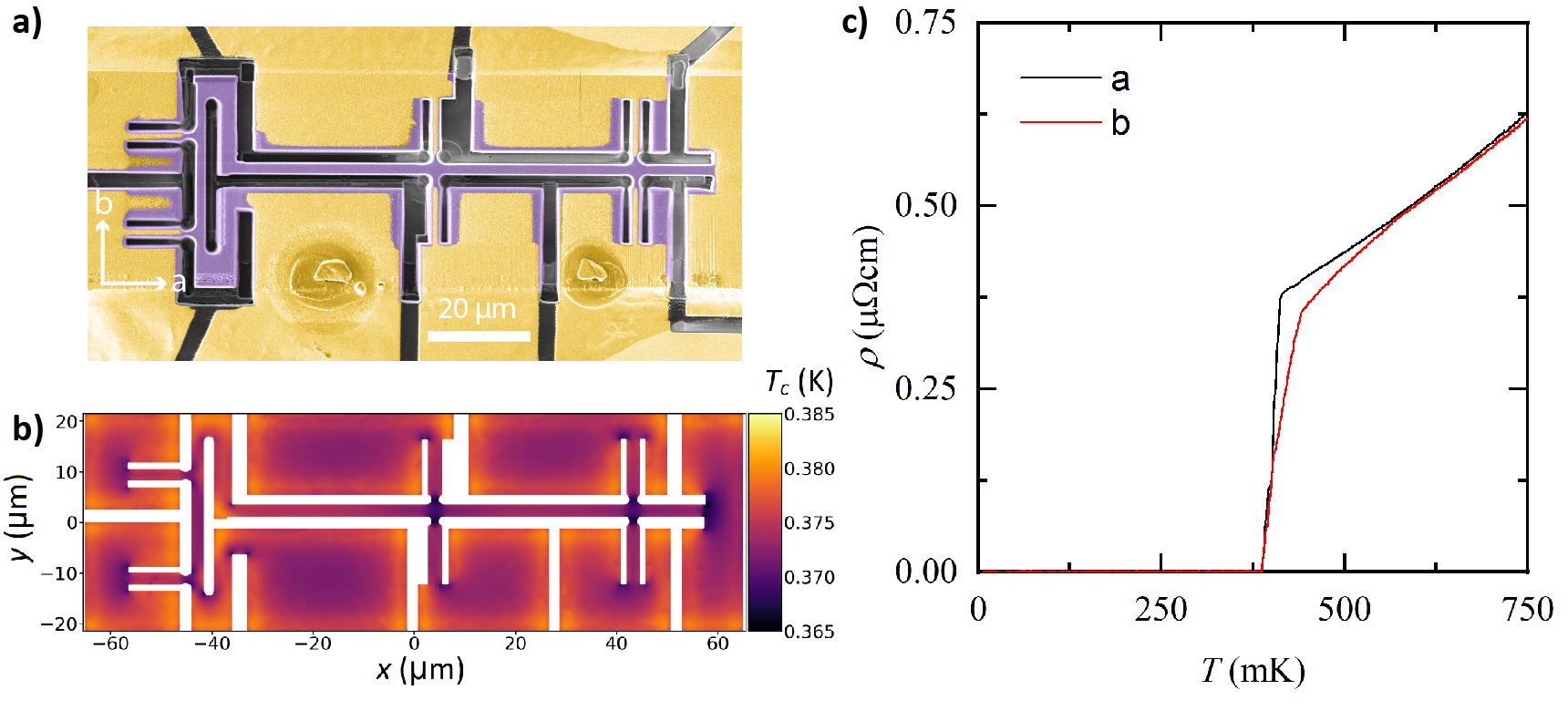}
\caption{(a) False-color SEM image of a strain-minimizing T-shape device of CeIrIn$_5$ on a CaF$_2$ substrate. (b) Simulated $T_c$ map of the device shown in (a). Variations in $T_c$ across the active sections of this device are limited to approximately 5~mK. (c) Resistivity measurements of the long ($a$) and short ($b$) legs of the device, showing the superconducting transition in agreement with a nearly unstrained device. For both legs, a state of zero resistance is reached at $T_c^0$=386~mK. The first onset of a superconducting transition takes place at $T_c^{\textrm{onset}}$=440~mK for the a-leg and 570~mK for the b-leg.}
\label{fig2}
\end{figure}

Fig.~\ref{fig2} summarizes the best attempt at such a low-strain device. Our simulations of this device show a maximal variation in $T_c$ of approximately 5~mK across the longer a-leg of the device and 10~mK across the shorter b-leg. Experimentally, we observe a sharp superconducting transition in the a-leg, at a temperature in good agreement with reports of bulk CeIrIn$_5$\cite{Petrovic2001,Chen2015a,Borth2002}. The transition measured over the b-leg, however, has a higher onset temperature of approximately 570~mK, but the resistance decreases slowly down to 450~mK before dropping down sharply. This intermediate region of slowly decreasing resistance is possibly related to a parallel conduction channel caused by redeposition of material removed by the FIB, while the sharp drop is intrinsic to CeIrIn$_5$. In this case, the geometry of the device makes the b-leg more prone to the formation of such parallel channels. The remaining difference between the two legs in the onset of the sharp transition (see Fig.~S4) can be attributed to an inhomogeneous local current flow pattern. For the $b$-leg of the device, the current flows around a corner with a higher $T_c$ at either end of the transport bar. This can affect the measured voltage at the contact points, leading to a reduced resistance in a certain temperature range. Above 570~mK and below 400~mK, the $a$ and $b$ resistivities overlap as expected and both reach zero at $T_c^0$=386~mK. 

\begin{figure}[ht]
\centering
\includegraphics[width=0.7\linewidth]{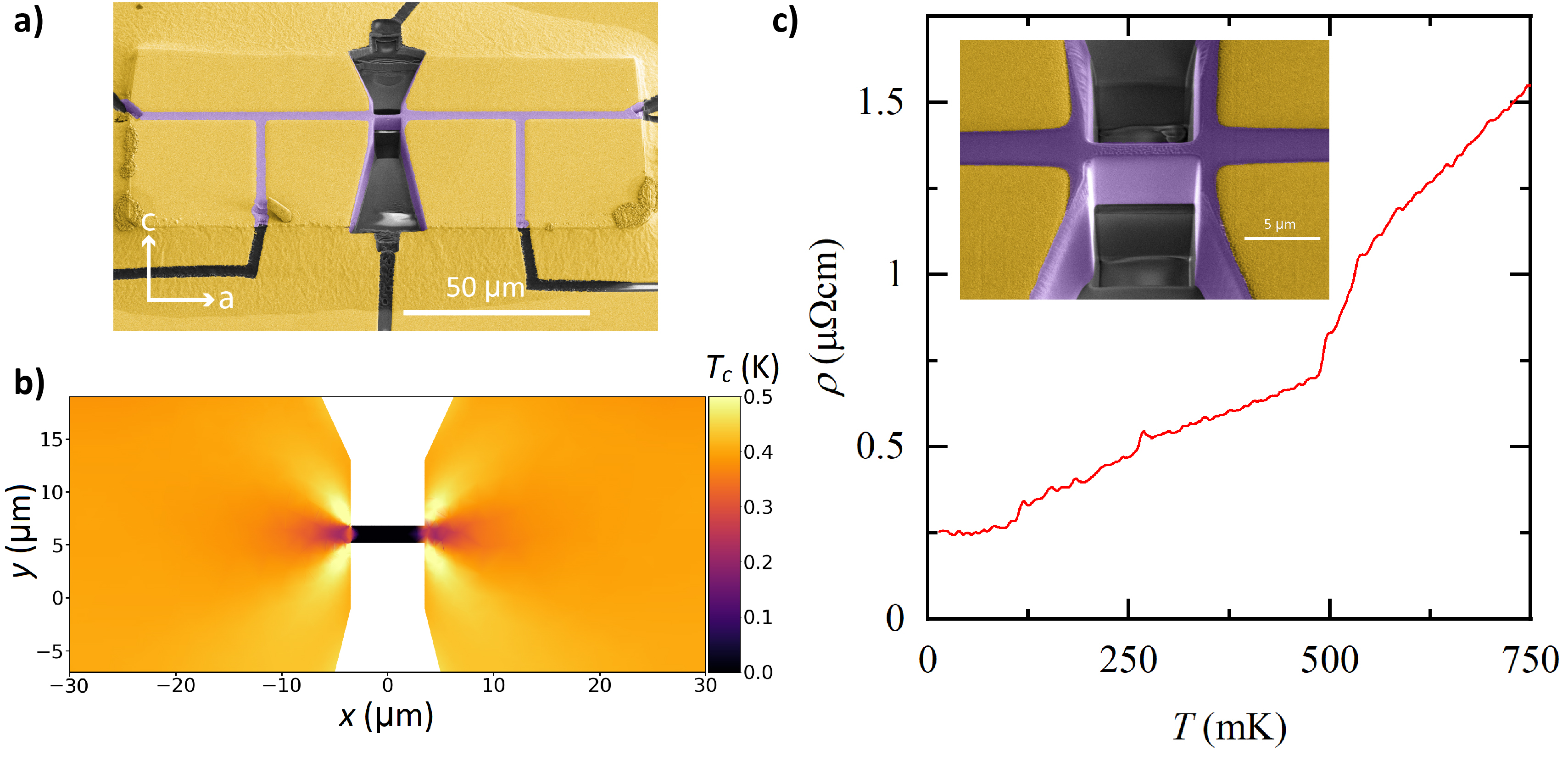}
\caption{(a) False-color SEM image of a CeIrIn$_5$ device on a fused silica substrate designed to maximize the strain in one section. (b) Simulated $T_c$ map of the device shown in (a). The bridge section in the center has $T_c$=0~K. (c) Resistivity measurement through the bridge section of the device (shown in the inset). The bridge does not become superconducting, but maintains a resistivity comparable to previous reports\cite{Bachmann2019} down to the base temperature of our experiment. }
\label{fig3}
\end{figure}

The elimination of strain demonstrated here paves the way towards finite size studies of confined heavy-fermion conductors without the additional worries of strain effects. Yet tailoring strong strain fields also enables novel functionality, such as intentionally creating sections of devices with enhanced or suppressed $T_c$. These approaches will be based on the high-strain substrate fused silica. The shape of the microstructure can focus the strain into a small area, which happens most effectively in a bridge-like structure: a thin bar between two large blocks of crystal. Our results for a device of this type are summarized in Fig.~\ref{fig3}. 

The device shown in Fig.~\ref{fig3}a has a 7~$\mu$m long and 2~$\mu$m wide bridge in which superconductivity is expected to be fully suppressed, as suggested by simulations of this device (see Fig.~\ref{fig3}b). Experimentally, this is indeed what we observe. In Fig.~\ref{fig3}c, we show the measured resistivity of this device as a function of temperature. Around 500~mK, a first transition takes place arising from the bulk slabs outside the bridge. Below this temperature, the resistivity gradually decreases only to end up around 0.25~$\mu\Omega$cm, a value in good agreement with previous reports in the normal state\cite{Bachmann2019}. The gradual decrease of resistivity is likely due to a combination of the usual metallic conductivity of CeIrIn$_5$ above $T_c$ and the region of partially suppressed superconductivity extending a few $\mu$m from either end of the bridge.

The structure shown in Fig.~\ref{fig3} is rather simple, and the geometric factors of length, width and thickness of the bridge dominantly set its $T_c$. Yet further, the depth and shape of the cuts made into the lamella on either side of the bridge play an important role as well as they are important in setting the elastic boundary condition for the problem. Cutting deep into the substrate, or cutting at a larger opening angle both increase the strain experienced by the bridge. Generally, direct contact between the bridge and the epoxy on the substrate is not beneficial and a stronger strain can be achieved by undercutting it. In devices with fully suppressed superconductivity, a compressive c-axis strain of approximately 0.6\% is typically the main driving factor, while the a-axis strain of about 0.5\% has an opposite sign between the in- and out-of-plane directions and therefore makes only a small contribution.  

\begin{figure}[ht]
\centering
\includegraphics[width=1\linewidth]{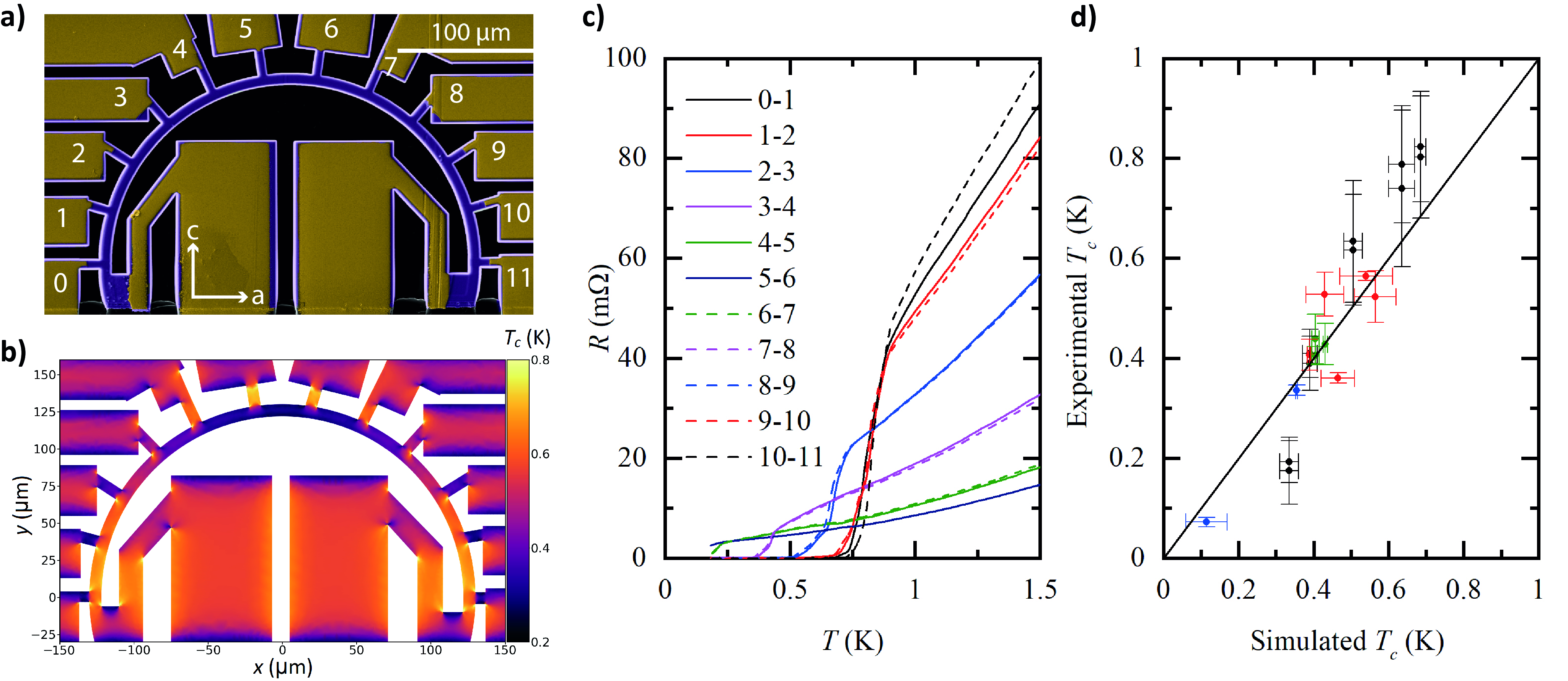}
\caption{(a) False-color SEM image of a circular device on a sapphire substrate, designed to exhibit a broad range of critical temperatures. The voltage contacts of the device are numbered 0 to 11. (b) Simulated $T_c$ map of the device shown in (a). (c) Experimental superconducting transitions measured in the device shown in (a). The numbers in the legend correspond to the pairs of contacts measured for the respective curves. (d) Experimental versus simulated critical temperatures for several devices. The straight line represents perfect agreement of simulation and experiment. Black dots were extracted from the semi-circular device, red and blue dots from the bulk and bridge sections of bridge-style devices respectively, and green dots from low-strain devices. Device sections with $T_c$=0~K are not included, as these typically have simulated critical temperatures slightly below zero, resulting from the very high strain. Vertical error bars are determined from the width of the measured transitions and horizontal error bars from the range of simulated $T_c$ values over the relevant device sections.}
\label{fig4}
\end{figure}

Having demonstrated both extremes of our technique, low strain and high strain, we now turn to the design of complex strain and $T_c$ patterns. This serves as a demonstration of the engineering capabilities of the approach, yet also tests our model over a wide range of temperatures. To this end, we have fabricated an approximately 4~$\mu$m thick semi-circular structure shown in Fig.~\ref{fig4}a. The direction of the strain relative to the crystallographic axes varies continuously along the circle, leading to either suppression or enhancement of $T_c$ in different sections. This is shown by our simulation, in Fig.~\ref{fig4}b.

In Fig.~\ref{fig4}c, we show the measured resistances between each of the pairs of adjacent contacts to the device. The results are in good agreement with our predictions and are symmetric between the left and right sides of the device, thereby confirming the accurate crystallographic alignment. As expected, the section between contacts 5 and 6 shows the lowest $T_c$, below the minimal temperature accessible to the measurement, and the sections 0-1 and 10-11 show the highest $T_c$, about twice the value of bulk CeIrIn$_5$. 

The overall agreement between our simulations and experiment is demonstrated by Fig.~\ref{fig4}d, where we have included experimental data and simulations from multiple sections of 8 devices. The good agreement confirms the validity of the model, and that reliable $T_c$ landscapes can be selectively written into superconductors using the FIB. The remaining discrepancies can be well explained by unintentional deviations of the real devices from the targeted models. For instance, the simulations assume that all devices are perfectly flat on the epoxy layer at exact crystallographic orientation, while in practice this alignment will be imperfect on the order of a percent. 

In conclusion, we have demonstrated microscopic control over the superconducting critical temperature $T_c$ of CeIrIn$_5$ over a range from 0~K to at least twice the bulk $T_c$, by making use of the thermal contraction of different substrate materials and by careful device design. We additionally demonstrate the validity of finite element simulations for the prediction of $T_c$, enabling the design of devices with particular $T_c$ landscapes as desired. With this, we have laid the groundwork for the design and fabrication of functional devices. The methodology presented here is entirely materials agnostic, and can be used to tailor strain fields in arbitrary materials on the micrometer scale. In the future, such exploitation of strain-induced phase transformations in strongly correlated materials will surely enable new types of spatially modulated structures.

\noindent\textbf{Acknowledgements}\\
M.R.v.D. acknowledges funding from the Rubicon research program with project number 019.191EN.010, which is financed by the Dutch Research Council (NWO). M.D.B. acknowledges EPSRC for PhD studentship support through grant number EP/L015110/1. Work at Los Alamos was carried out under the auspices of the U.S. Department of Energy, Office of Science, Basic Energy Sciences, Materials Sciences and Engineering Division.

\noindent\textbf{Data availability}\\
The data that support the findings of this study are openly available in the Zenodo repository at https://doi.org/10.5281/zenodo.5785684.

\end{document}